\documentclass[twocolumn]{aastex7}

\begin{document}

\title{High Proper Motion Discoveries from the UKIRT Hemisphere Survey}
\author[0009-0003-4167-5207]{Wings Zhang}
\affiliation{United States Naval Observatory, Flagstaff Station, 10391 West Naval Observatory Rd., Flagstaff, AZ 86005, USA}
\email{}

\author[0000-0002-6294-5937]{Adam C. Schneider}
\affil{United States Naval Observatory, Flagstaff Station, 10391 West Naval Observatory Rd., Flagstaff, AZ 86005, USA}
\email{}

\author[0000-0003-2235-761X]{Thomas P. Bickle}
\email{}
\affil {School of Physical Sciences, The Open University, Milton Keynes, MK7 6AA, UK}
\affil {Backyard Worlds: Planet 9}

\author[0000-0002-6523-9536]{Adam J. Burgasser}
\email{}
\affil{Center for Astrophysics and Space Science, University of California San Diego, La Jolla, CA 92093, USA}

\author[0000-0002-1420-1837]{Emma Softich}
\email{}
\affiliation{Department of Astronomy \& Astrophysics, UC San Diego, La Jolla, CA, USA}

\author[0000-0001-7519-1700]{Federico Marocco}
\email{}
\affil{IPAC, Mail Code 100-22, Caltech, 1200 E. California Blvd., Pasadena, CA 91125, USA}

\author[0000-0001-8170-7072]{Daniella Bardalez Gagliuffi}
\email{}
\affil{Department of Physics \& Astronomy, Amherst College, 25 East Drive, Amherst, MA 01003, USA}

\author[0000-0001-6251-0573]{Jacqueline K. Faherty}
\email{}
\affil{Department of Astrophysics, American Museum of Natural History, Central Park West at 79th St., New York, NY 10024, USA}

\author[0000-0002-1125-7384]{Aaron M. Meisner}
\email{}
\affil{NSF's National Optical-Infrared Astronomy Research Laboratory, 950 N. Cherry Ave., Tucson, AZ 85719, USA}

\author[0000-0003-4269-260X]{J. Davy Kirkpatrick}
\email{}
\affil{IPAC, Mail Code 100-22, Caltech, 1200 E. California Blvd., Pasadena, CA 91125, USA}

\author[0000-0002-2387-5489]{Marc J. Kuchner}
\email{}
\affil{Exoplanets and Stellar Astrophysics Laboratory, NASA Goddard Space Flight Center, 8800 Greenbelt Road, Greenbelt, MD 20771, USA}

\author[0000-0003-4905-1370]{Martin Kabatnik}
\email{}
\affil{Backyard Worlds: Planet 9}

\author[0000-0001-8662-1622]{Frank Kiwy}
\email{}
\affil{Backyard Worlds: Planet 9}

\author[0000-0003-4864-5484]{Arttu Sainio}
\email{}
\affil{Backyard Worlds: Planet 9}

\author[0000-0002-7587-7195]{J{\"o}rg	Sch{\"u}mann}
\email{}
\affil{Backyard Worlds: Planet 9}

\author{Karl Selg-Mann}
\email{}
\affil{Backyard Worlds: Planet 9}

\author[0000-0003-4714-3829]{Nikolaj Stevnbak Andersen}
\email{}
\affil{Backyard Worlds: Planet 9}

\author[0000-0001-8731-9281]{Bruce Baller}
\email{}
\affil{Backyard Worlds: Planet 9}

\author{Paul Beaulieu}
\email{}
\affil{Backyard Worlds: Planet 9}

\author{John Bell}
\email{}
\affil{Backyard Worlds: Planet 9}

\author[0000-0001-7896-5791]{Dan Caselden}
\email{}
\affiliation{Department of Astrophysics, American Museum of Natural History, Central Park West at 79th Street, NY 10024, USA}

\author[0000-0002-7630-1243]{Guillaume Colin}
\email{}
\affil{Backyard Worlds: Planet 9}

\author{Alexandru Dereveanco}
\email{}
\affil{Backyard Worlds: Planet 9}

\author{Christoph Frank}
\email{}
\affil{Backyard Worlds: Planet 9}

\author{Konstantin Glebov}
\email{}
\affil{Backyard Worlds: Planet 9}

\author[0000-0002-8960-4964]{L\'eopold Gramaize}
\email{}
\affil{Backyard Worlds: Planet 9}

\author[0000-0002-7389-2092]{Leslie K. Hamlet}
\email{}
\affil{Backyard Worlds: Planet 9}

\author{David W. Martin}
\email{}
\affil{Backyard Worlds: Planet 9}

\author{William Pendrill}
\email{}
\affil{Backyard Worlds: Planet 9}

\author{St{\'e}phane Perlin}
\email{}
\affil{Backyard Worlds: Planet 9}

\author{Andres Stenner}
\email{}
\affil{Backyard Worlds: Planet 9}

\author{Christopher Tanner}
\email{}
\affil{Backyard Worlds: Planet 9}

\author[0000-0001-5284-9231]{Melina Th{\'e}venot}
\email{}
\affil{Backyard Worlds: Planet 9}

\author{Vinod Thakur}
\email{}
\affil{Backyard Worlds: Planet 9}

\begin{abstract}

We used the third data release of the UKIRT Hemisphere Survey to locate previously unrecognized high proper motion objects. We identify a total of 127 new discoveries with total proper motions $\gtrsim$300 mas yr$^{-1}$. A significant fraction of these sources with counterparts in the Gaia DR3 catalog are found to be distant ($>$100 pc) low-mass stars, where their large tangential velocities and placement on color-magnitude diagrams suggest that they are likely low-metallicity M-type subdwarfs. Optical spectroscopy is used to confirm the low-mass and low-metallicity for two such sources. Using available optical and infrared photometry, we estimate the spectral type for all non-Gaia sources and find 10 likely late-M dwarfs, 15 objects with colors most consistent with L-type dwarfs, and 9 possible T-type dwarfs. Follow-up spectroscopy is needed to confirm spectral types and further characterize these new discoveries.

\end{abstract}


\section{Introduction} 
The UKIRT Hemisphere Survey (UHS) is a near-infrared survey using the United Kingdom Infrared Telescope (UKIRT) to observe $\approx$12,700 square degrees of the Northern hemisphere \citep{dye2018}. The third UHS data release (DR3; \citealt{schneider2025}) contains the $H$-band portion of the survey and includes proper motion measurements for those sources matched at more than one band ($J$, $H$, and/or $K$) during the band-merging process. Proper motions have historically been shown to be exceptionally valuable for identifying a variety of astrophysically interesting sources, including our nearest neighbors (e.g., \citealt{innes1915, barnard1916, luhman2013, luhman2014, kirkpatrick2024}) and low-metallicity galactic oddballs (e.g., \citealt{schweitzer1999, scholz2004, burgasser2006, kirkpatrick2021, meisner2021}). 

We have searched the UHS DR3 catalog for previously unknown high proper motion sources. We describe the methodologies and process of the selection process in Section \ref{sec:targets}, the spectroscopic follow-up observations of two targets in Section \ref{sec:obs}, and our analysis of newly discovered systems in Section \ref{sec:analysis}. Finally, we conclude our findings in Section \ref{sec:conclusion}.

\section{Candidate Selection}
\label{sec:targets}

All UHS data are processed by the Wide Field Astronomy Unit (WFAU) at the University of Edinburgh. The pipeline used to produce the UHS DR3 source catalog determines proper motions using the method described in \cite{collins2012}. We select all objects in UHS DR3 with total proper motion measurements $>$300 mas yr$^{-1}$ as candidates. This proper motion threshold was set in order to select a practical number of candidates to inspect individually. We further required the jErrBits, hErrBits, kErrBits flags, which are WFAU post-processing error bits, to be zero to be considered candidates. Setting these flags to zero avoids selecting sources that have been affected by saturated pixels, blends, bad flat fields, bad pixels, and crosstalk. We also implemented an uncertainty threshold of $\leq$20 mas yr$^{-1}$ for each proper motion component. To determine this uncertainty threshold, we cross-matched the list of known L, T, and Y dwarfs in the UHS footprint from \cite{schneider2023} with the UHS DR3 catalog, finding 776 matches. We found that 765 of the 776 objects found with UHS DR3 proper motions have proper motion uncertainties lower than 10 mas yr$^{-1}$ and 773 objects with proper motion uncertainties lower than 20 mas yr$^{-1}$. We opted for a larger proper motion uncertainty threshold of 20 mas yr$^{-1}$ in an attempt to detect fainter moving objects than those already known. These selection criteria resulted in 24,391 total candidates, which were next individually inspected to confirm proper motions.

\begin{figure*}
\plotone{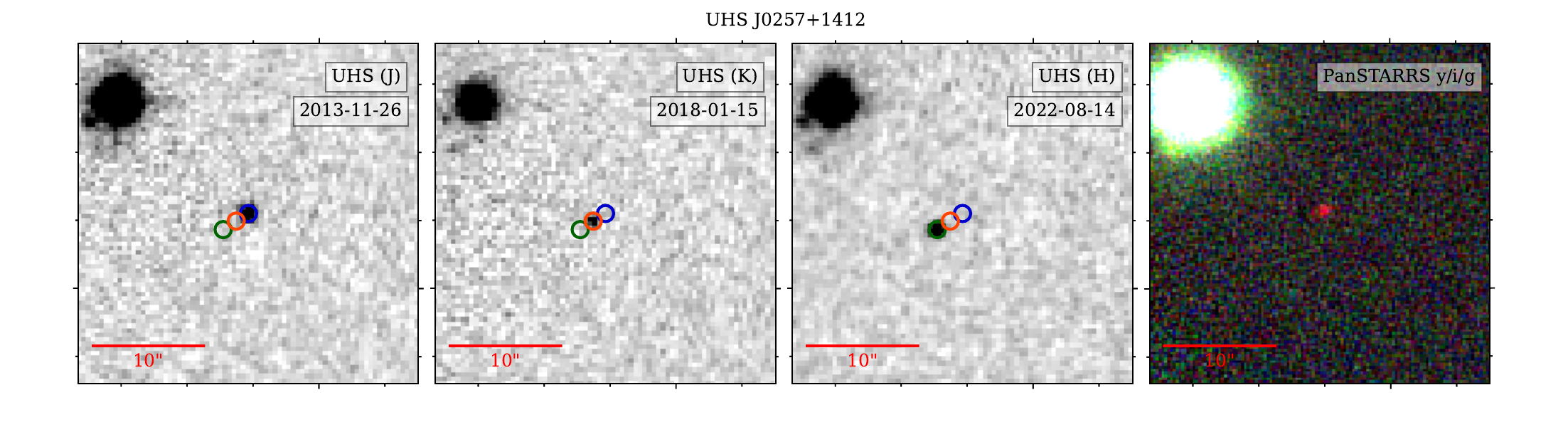}
\caption{Example finder chart of a new high proper motion discovery identified in this work. The circles in the left three panels indicate the measured $J$-band position (blue), $H$-band position (green), and $K$-band position (red). Note the red color in the Pan-STARRS image, indicating a very red optical color and therefore a likely cold effective temperature.}
\label{fig:finder}
\end{figure*}

Each candidate was visually inspected by two of us (WZ and AS) using UHS and Pan-STARRS \citep{chambers2016} images. We included Pan-STARRS in addition to UHS to help confirm sources where UHS images were not conclusive, as Pan-STARRS images have the advantage of highlighting exceptionally red-optical sources, which could indicate a low-temperature and/or substellar nature. Figure \ref{fig:finder} shows an example finder chart with UHS $J$-, $H$-, and $K$-band images in addition to a Pan-STARRS three-color image. Common reasons for rejecting candidates during the visual inspection process includes instrumental artifacts, diffraction spikes, and blending in crowded fields. After the visual inspection vetting, the number of remaining candidates was 1,848.

We then checked each candidate against SIMBAD \citep{wenger2000} and only kept those objects that did not have SIMBAD entries, or those with SIMBAD entries but zero references. We found that the majority of the motion-confirmed objects had been discovered previously, though 239 candidates still remained after this process. 

After the SIMBAD cross-match, we performed one additional cross-check with the Backyard Worlds: Planet 9 Citizen Science Project (BYW; \citealt{kuchner2017}). BYW has similar aims as the UHS-based survey presented in this work: namely, to identify previously unknown high-proper motion infrared sources. While this work uses UHS to identify sources, BYW uses unWISE images \citep{lang2014, meisner2019} from the Wide-field Infrared Survey Explorer \citep{wright2010}, which have a pixel scale of 2\farcs75 per pixel at infrared bands centered at 3.4 $\mu$m and 4.6 $\mu$m. Since these projects have similar objectives, it is not surprising that the discovery list from this work has some overlap with the objects discovered by citizen scientists from BYW. We removed any candidates from our list that were already being prepared for publication by the BYW team, leaving 127 motion-confirmed discoveries from the UHS survey. The properties of these sources, including UHS proper motions and photometry, are given in Table \ref{tab:props}. We recognize the contributions of the citizen scientists from the BYW project who independently identified many of the discoveries in this work in Table \ref{tab:props}. We note that the high proper motions of these sources caused the merging step in building the UHS DR3 merged catalog to not always include all detections of each object (e.g., an object could have a proper motion determined from its J- and K-band detections, but an H-band detection for this object exists on a separate line in the merged catalog). We therefore include all photometric measurements of each source from the DR3 catalog for completeness. 

After these checks, we noted the work of \cite{karpov2025}, which similarly focused on uncovering new high–proper motion objects using WISE, with many of its discoveries not yet migrated into SIMBAD. We found a total of seven objects from \cite{karpov2025} in common with our list of 127 discoveries. We retain these objects as independently discovered sources for our analysis, but note the original discovery reference for these seven sources (UHS J0023+4353, UHS J0043+2754, UHS J1101+4307, UHS J1326+5543, UHS J1443+3119, UHS J1736+0543, and UHS J2216+3136) should be \cite{karpov2025}.

\startlongtable
\begin{deluxetable*}{llccccccc}
\label{tab:props}
\tablecaption{UHS High Proper Motion Discovery Properties }
\tablehead{
\colhead{Column Label} & \colhead{Description} & \colhead{Example} & Units \\}
\startdata
UHS-Name & UHS DR3 name & UHS J000904.28+330629.3 & \dots \\
RAdeg-UHS & UHS DR3 right ascension & 2.2678522 & degrees \\
DEdeg-UHS & UHS DR3 declination & 33.1081297 & degrees \\
Epoch & Epoch of UHS DR3 position & 2019.753 & years \\
UHS-pmRA & UHS $\mu$$_{\alpha}$ & 435.85 & mas yr$^{-1}$ \\
e\_UHS-pmRA & UHS uncertainty of $\mu$$_{\alpha}$ & 9.30 & mas yr$^{-1}$ \\
UHS-pmDE & UHS $\mu$$_{\delta}$ & 16.794 & mas yr$^{-1}$ \\ 
e\_UHS-pmDE & UHS uncertainty of $\mu$$_{\delta}$ & 9.30 & mas yr$^{-1}$ \\
Jmag & UHS J magnitude & 16.896 & mag \\
e\_Jmag & Uncertainty of UHS J magnitude & 0.020 & mag \\
Hmag & UHS H magnitude & 16.513 & mag \\
e\_Hmag & Uncertainty of UHS H magnitude & 0.024 & mag \\
Kmag & UHS K magnitude & 16.124 & mag \\
e\_Kmag & Uncertainty of UHS K magnitude & 0.043 & mag \\
BYW-discoverer-code\tablenotemark{a} & BYW discoverer code & Q,T,X & \dots \\
Gaia-ID & Gaia DR3 ID & Gaia DR3 2874116561714197120 & \dots \\
Gaia-plx & Parallax from Gaia DR3 & 4.50 & mas \\
e\_Gaia-plx & Uncertainty of Gaia DR3 parallax & 0.74 & mas \\
Gaia-pmRA & Gaia DR3 $\mu$$_{\alpha}$ & 435.12 & mas yr$^{-1}$ \\
e\_Gaia-pmRA & Gaia DR3 uncertainty of $\mu$$_{\alpha}$ & 0.82 & mas yr$^{-1}$ \\
Gaia-pmDE & Gaia DR3 $\mu$$_{\delta}$ & 6.24 & mas yr$^{-1}$ \\ 
e\_Gaia-pmDE & Gaia DR3 uncertainty of $\mu$$_{\delta}$ & 0.54 & mas yr$^{-1}$ \\
Gmag & Gaia DR3 G magnitude & 20.329 & mag \\
e\_Gmag & Uncertainty of Gaia DR3 G magnitude & 0.006 & mag \\
BPmag & Gaia DR3 G$_{\rm BP}$ magnitude & 21.345 & mag \\
e\_BPmag & Uncertainty of Gaia DR3 G$_{\rm BP}$ magnitude & 0.160 & mag \\
RPmag & Gaia DR3 G$_{\rm RP}$ magnitude & 18.991 & mag \\
e\_RPmag & Uncertainty of Gaia DR3 G$_{\rm RP}$ magnitude & 0.049 & mag \\
Vtan & Tangential velocity & 458 & km s$^{-1}$ \\
e\_Vtan & Uncertainty of tangential velocity & 75 & km s$^{-1}$ \\
Name-PS1 & Pan-Starrs DR2 designation & PSO J002.2672+33.1081 & \dots \\
gmag & Pan-Starrs $g$ magnitude & \dots & mag \\
e\_gmag & Pan-Starrs $g$ uncertainty & \dots & mag \\
rmag & Pan-Starrs $r$ magnitude & 21.422 & mag \\
e\_rmag & Pan-Starrs $r$ uncertainty & 0.136 & mag \\
imag & Pan-Starrs $i$ magnitude & 19.527 & mag \\
e\_imag & Pan-Starrs $i$ uncertainty & 0.022 & mag \\
zmag & Pan-Starrs $z$ magnitude & 18.622 & mag \\
e\_zmag & Pan-Starrs $z$ uncertainty & 0.034 & mag \\
ymag & Pan-Starrs $y$ magnitude & 18.247 & mag \\
e\_ymag & Pan-Starrs $y$ uncertainty & 0.046 & mag \\
Name-CWISE & CatWISE2020 designation & J000904.18+330628.9 & \dots \\
W1mag & CatWISE2020 W1 magnitude & 15.747 & mag \\
e\_W1mag & CatWISE2020 W1 magnitude uncertainty & 0.023 & mag \\
W2mag & CatWISE2020 W2 magnitude & 15.639 & mag \\
e\_W2mag & CatWISE2020 W2 magnitude uncertainty & 0.045 & mag \\
Note & Note\tablenotemark{b} & \dots & \dots \\
\enddata
\tablecomments{This table is available in its entirety in a machine-readable form in the online journal.}
\tablenotetext{a}{BYW Discoverer Codes:
A = Nikolaj Stevnbak Andersen,
B = Bruce Baller,
C = Paul Beaulieu,
D = John Bell,
E = Tom Bickle,
F = Dan Caselden,
G = Guillaume Colin,
H = Alexandru Dereveanco,
I = Christoph Franck,
J = Konstantin Glebov,
K = Sam Goodman,
L = L{\'e}opold Gramaize,
M = Les Hamlet,
N = Martin Kabatnik,
O = Frank Kiwy,
P = David W. Martin,
Q = William Pendrill,
R = St{\'e}phane Perlin,
S = Arttu Sainio,
T = J{\"o}rg Sch{\"u}mann,
U = Karl Selg-Mann,
V = Andres Guillermo Stenner,
W = Christopher Tanner,
X = Melina Th{\'e}venot,
Y = Vinod Thakur}
\tablenotetext{b}{Notes: 1 = WISE data from AllWISE Source Catalog; 2 = WISE data from CatWISE2020 Reject Catalog}
\end{deluxetable*}

\section{Observations}
\label{sec:obs}
\subsection{Lick/KAST}
We obtained follow-up optical spectra for UHS J1803+1516 and UHS J1855+4054 with the KAST double spectrograph \citep{miller1994} on the 3 m Shane Telescope located at the Lick Observatory on Mount Hamilton, California. Data were acquired on July 27th, 2025. Observations were carried out with the 0.5\arcsec\ slit for J1803+1516 and 1.5\arcsec\ slit for J1855+4054. The 600/7500 red grating was used, which provided resolutions of $\lambda / \Delta\lambda \approx 1800$ over the 6300-9000 \r{A} wavelength rage. For both targets, the total exposure time was 1800 seconds. Data were reduced using the \texttt{kastredux}\footnote{https://github.com/aburgasser/kastredux} package. 

\section{Analysis}
\label{sec:analysis}
\subsection{Proper Motions}
\begin{figure}
\plotone{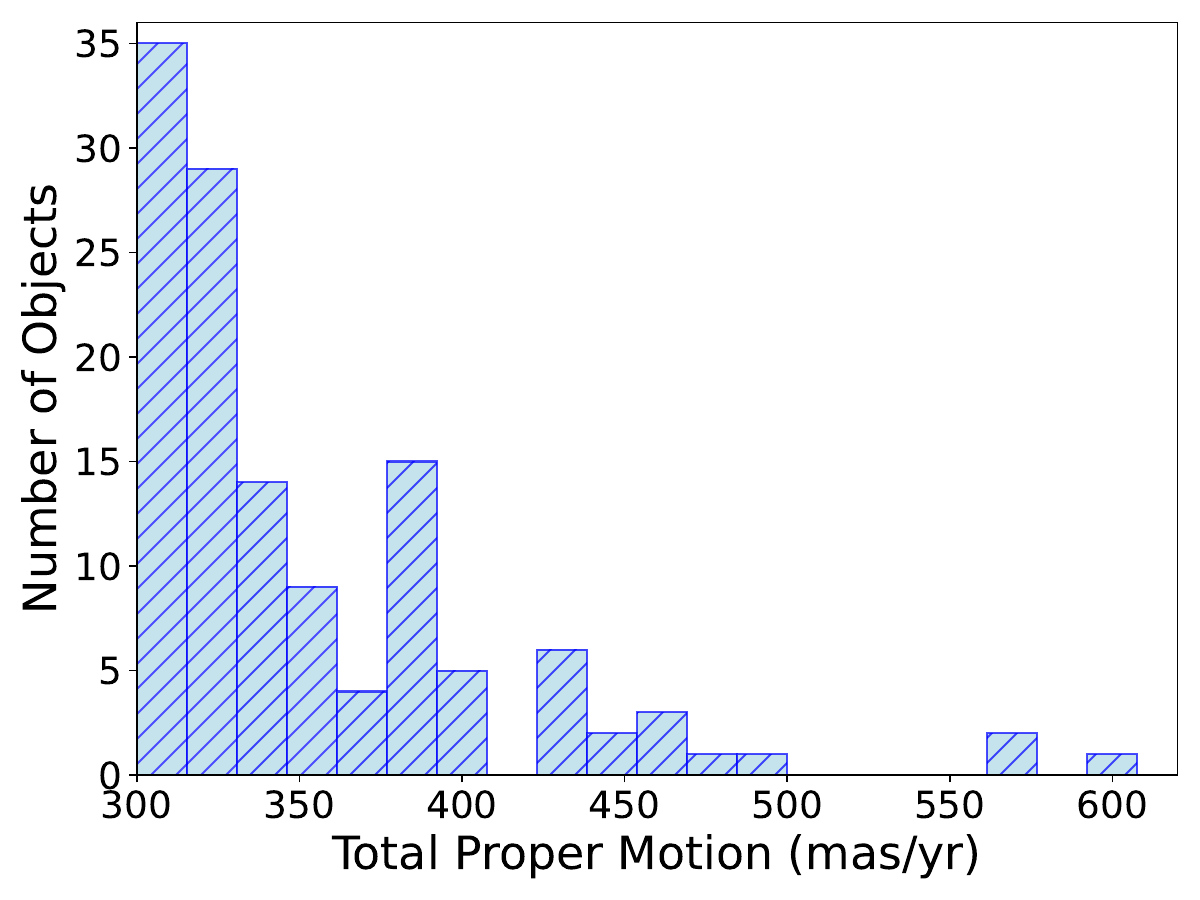}
\caption{Histogram of total proper motion values for new discoveries in this work.}
\label{fig:pm}
\end{figure}

One-hundred and twenty-seven new sources were found with total proper motions $>$300 mas yr$^{-1}$. Three of our newly discovered objects have total proper motions $>$500 mas yr$^{-1}$ (UHS J0244+3011, UHS J0609+3539, and UHS J1557+1212), with UHS J0244+3011 having the largest total proper motion in the sample of $\sim$600 mas yr$^{-1}$. Figure \ref{fig:pm} shows a histogram of the total proper motion values for the sample of new objects. The largest proper motions detectable in the UHS DR3 catalog depends on the time baseline between $J$-, $H$-, and/or $K$-band observations \citep{schneider2025}, as well as the 2\arcsec\ matching radius used in the catalog merging pipeline. The largest proper motion source identified in our search was LP 440-52 \citep{luyten1979}, an M7 subdwarf \citep{kesseli2019} with a total proper motion of 1220 mas yr$^{-1}$ \citep{gaia2021}. Therefore objects with proper motions larger than $\sim$1200 mas yr$^{-1}$ are likely unmatched in UHS DR3, and were naturally excluded as part of this survey.

\subsection{Photometry}
To aid in characterizing all of the discoveries in this work, we gathered photometry from the Gaia DR3 \citep{gaia2021}, Pan-STARRS DR2 \citep{chambers2016}, and the CatWISE2020 \citep{marocco2021} catalogs, using a 5\arcsec\ search radius. We compared proper motion and photometry values to ensure cross-matches were consistent with the objects in our sample. Photometry for each source can be found in Table \ref{tab:props}.

\subsection{Gaia Sources}
\label{sec:gaia}
\begin{figure*}
\plotone{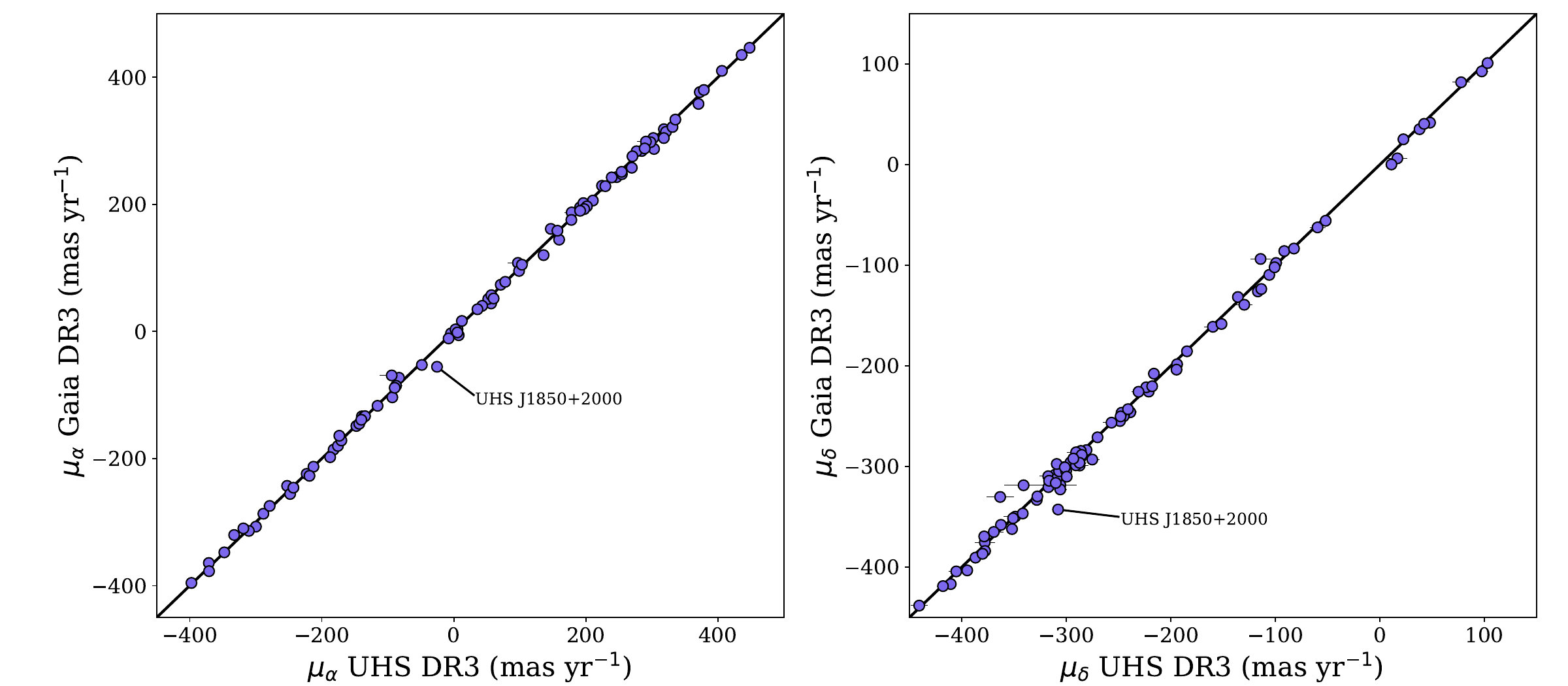}
\caption{A comparison of the measured proper motion components from UHS DR3 and Gaia DR3. The solid black line indicates a ratio of unity. }
\label{fig:pmcomp}
\end{figure*}

Of the 127 new proper motion discoveries, 94 have entries in the Gaia DR3 catalog \citep{gaia2021}. Three of these sources (UHS J0423+0016, UHS J0440+0831, UHS J1413+3121) only have positions and photometry, but no proper motion or parallax measurements. Table \ref{tab:props} also includes the Gaia DR3 astrometry and photometry of each new object with entries in the Gaia DR3 catalog. Of the 94 objects with proper motion measurements in UHS DR3 and Gaia DR3, the vast majority (90 objects) have proper motion components consistent within 3$\sigma$ (Figure \ref{fig:pmcomp}). Only one object (UHS J1850+2000) has $\mu_{\alpha}$ and $\mu_{\delta}$ components that differ by more than 4$\sigma$. A closer inspection of the UHS images of UHS J1850+2000 reveals a relatively crowded field around this source where it is blended with an unrelated background source that likely affected its measured UHS proper motion, leading to $\mu_{\alpha}$ and $\mu_{\delta}$ differences of $\sim$9$\sigma$ and $\sim$10$\sigma$, respectively. 

Tangential velocities ($V_{\rm tan}$) have been shown to be useful for differentiating between different Galactic components (e.g., thin disk, thick disk, and halo). To characterize the Gaia sample, we calculated $V_{\rm tan}$ values for each object using the Gaia proper motions and parallaxes. All Gaia sources with measured parallaxes were found to have large tangential velocities ($V_{\rm tan}$ $\gtrsim$ 190 km s$^{-1}$). We compared our sources to the criteria outlined in \cite{dupuy2012}, where they found that objects with $V_{\rm tan}$ $>$ 91+56$e^{(0.024{\rm e}V_{\rm tan})}$ and 1 $<$ e$V_{\rm tan}$ $<$ 70 have a probability of belonging to the thin disk of $<$10\%. Of the 80 objects that have a $V_{\rm tan}$ uncertainty (e$V_{\rm tan}$) within the range listed above, all objects satisfy this criteria and are therefore unlikely to be thin disk members. This is consistent with the criteria described in \cite{nissen2004}, where it was determined that most objects with total velocities greater than 180 km s$^{-1}$ are likely halo members. 

\begin{figure*}
\plotone{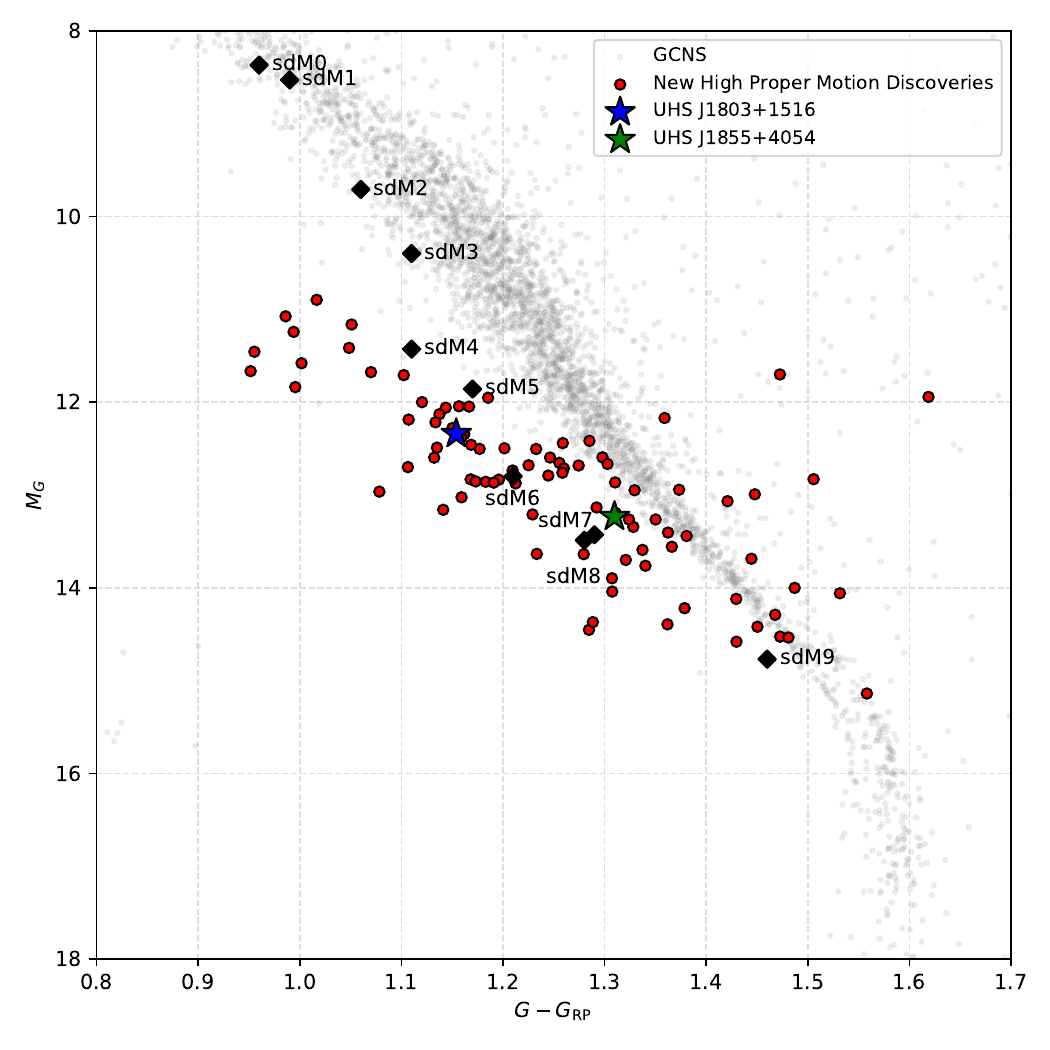}
\caption{Color-magnitude diagram comparing the newly discovered high proper motion objects (red) with those from the Gaia Catalog of Nearby Stars (GCNS; gray). Also shown are the median subdwarf colors for sdM0 through sdM9 spectral types (black diamonds) as described in \ref{sec:gaia}. The two objects with spectra, as shown in \ref{fig:spectra}, are highlighted with a blue or green star.}
\label{fig:cmd}
\end{figure*}

To further explore the nature of the sample of Gaia-matched discoveries, we created a color magnitude diagram (Figure \ref{fig:cmd}) to compare the positions of these objects to field objects from the Gaia Catalog of Nearby Stars (GCNS; \citealt{smart2021}). For this comparison, we used all GCNS members with a distance measurement within 25 pc. We did not correct photometry for interstellar reddening, as our objects are generally located outside of the galactic plane (because UHS avoids areas covered by previous UKIDSS surveys, including the Galactic Plane Survey), and reddening at Gaia wavelengths is generally negligible within a few hundred pc outside of the plane (e.g., \citealt{wang2025}). The figure shows that most of the objects in this sample lie below and towards the blue side of the empirical main sequence outlined by GCNS objects. This follows the patterns seen for spectroscopically confirmed subdwarfs (e.g., \citealt{hejazi2020, zhang2021}), and supports the supposition that many of these objects are likely old, low-metallicity members of the thick disk or halo. 

For the Gaia sample, we use the measured parallaxes and Gaia photometry to estimate approximate spectral types. While typical Gaia colors and absolute magnitudes for normal M-type dwarfs exist in the literature (e.g., \citealt{kiman2019}), the determination that most of this sample are likely low-metallicity subdwarfs makes comparisons to main-sequence M dwarfs unsuitable. We therefore created a subdwarf sequence of Gaia absolute magnitudes. We used the list of M-type subdwarfs provided in \cite{hejazi2022} to establish a reference list of subdwarfs with known spectral types. While the \cite{hejazi2022} sample is well populated from the sdM0 to sdM6 spectral types, there are very few objects with spectral types later than $\sim$sdM6.5. We therefore supplemented this list of objects with late M-type subdwarfs from \cite{kirkpatrick2014}, \cite{kirkpatrick2016}, \cite{lodieu2019}, and \cite{zhang2021}. We established a list of 333 objects with known spectral types from sdM0 to sdM9. The photometry was compiled to create a subdwarf sequence of median absolute magnitudes and colors. We show the $M_{G}$ vs.\ $G-G_{RP}$ sequence in Figure \ref{fig:cmd}, and provide median values for absolute Gaia photometry in Table \ref{tab:sds}. The listed uncertainties are the 16th and 84th percentile ranges. Using the Gaia parallaxes and apparent magnitudes from Table \ref{tab:props}, we calculated absolute magnitudes for each of the 91 Gaia candidates. These were then interpolated with the subdwarf sequence to determine approximate spectral types. The resulting spectral types for the different magnitudes were averaged to determine final spectral type estimates rounded to the nearest half-type, which are given in Table \ref{tab:g_spt}. The earliest spectral type estimate found for the Gaia sample is sdM3.5 (UHS J0244+3011), while the latest type is sdM9 (UHS J1517+5553 and UHS J2135+1825). 

\begin{deluxetable*}{ccccccccc}
\label{tab:sds}
\tablecaption{Median Gaia Absolute Magnitudes of M-type Subdwarfs}
\tablehead{
\colhead{Spectral} & \colhead{$M_{\rm G}$} & \colhead{$M_{\rm BP}$} & \colhead{$M_{\rm RP}$} & \colhead{N} \\
\colhead{Type} & \colhead{(mag)} & \colhead{(mag)} & \colhead{(mag)} & \colhead{} }
\startdata
sdM0-sdM0.5 & 8.37$^{+0.33}_{-0.56}$ & 9.31$^{+0.34}_{-0.56}$ & 7.41$^{+0.33}_{-0.56}$ & 21 \\
sdM1-sdM1.5 & 8.53$^{+0.44}_{-1.00}$ & 9.52$^{+0.53}_{-1.00}$ & 7.54$^{+0.43}_{-1.01}$ & 45 \\
sdM2-sdM2.5 & 9.71$^{+1.01}_{-0.66}$ & 10.75$^{+0.98}_{-0.80}$ & 8.65$^{+0.99}_{-0.67}$ & 51 \\
sdM3-sdM3.5 & 10.40$^{+0.91}_{-0.51}$ & 11.68$^{+0.96}_{-0.53}$ & 9.29$^{+0.95}_{-0.53}$ & 52 \\
sdM4-sdM4.5 & 11.43$^{+1.36}_{-0.29}$ & 12.72$^{+1.45}_{-0.40}$ & 10.32$^{+1.26}_{-0.26}$ & 26 \\
sdM5-sdM5.5 & 11.86$^{+0.67}_{-0.36}$ & 13.32$^{+0.82}_{-0.43}$ & 10.69$^{+0.61}_{-0.37}$ & 27 \\
sdM6-sdM6.5 & 12.80$^{+0.79}_{-0.42}$ & 14.21$^{+0.71}_{-0.63}$ & 11.59$^{+0.77}_{-0.36}$ & 71 \\
sdM7-sdM7.5 & 13.43$^{+0.64}_{-0.58}$ & 14.72$^{+0.64}_{-1.02}$ & 12.14$^{+0.52}_{-0.44}$ & 23 \\
sdM8-sdM8.5 & 13.49$^{+0.22}_{-0.37}$ & 15.30$^{+0.64}_{-0.46}$ & 12.21$^{+0.17}_{-0.41}$ & 10 \\
sdM9-sdM9.5 & 14.77$^{+0.12}_{-1.32}$ & 16.53$^{+0.91}_{-1.13}$ & 13.31$^{+0.08}_{-1.21}$ & 7 \\
\enddata
\tablecomments{This table gives median absolute magnitude values for Gaia photometry of M-type subdwarfs, where uncertainties are the 16 and 84 percentile ranges and N is the number of objects per spectral type bin. }
\end{deluxetable*}

\begin{deluxetable}{cccc}
\tabletypesize{\scriptsize}
\tablecaption{Spectral Type Estimates for Gaia Sources}
\label{tab:g_spt}
\tablehead{\colhead{UHS} & \colhead{Spectral} & \colhead{UHS} & \colhead{Spectral} \\
\colhead{Name} & \colhead{Type\tablenotemark{*}} & \colhead{Name} & \colhead{Type\tablenotemark{*}}}
\startdata
UHS J0009+3306 & [M7.5] & UHS J0916+1648 & [M6.5] \\
UHS J0017+4345 & [M7.5] & UHS J0929+3903 & [M8.5] \\
UHS J0039+2750 & [M8] & UHS J0956+1621 & [M4] \\
UHS J0053+3738 & [M8] & UHS J1006+1510 & [M6] \\
UHS J0105+4310 & [M5] & UHS J1018+3532 & [M5.5] \\
UHS J0139+4059 & [M5.5] & UHS J1039+3632 & [M6.5] \\
UHS J0143+2514 & [M8] & UHS J1106+1617 & [M6] \\
UHS J0222+3519 & [M5.5] & UHS J1108+3510 & [M6] \\
UHS J0237+4738 & [M5] & UHS J1118+3000 & [M7.5] \\
UHS J0243+0859 & [M8.5] & UHS J1121+3305 & [M8.5] \\
UHS J0244+3011 & [M3.5] & UHS J1139+3646 & [M6.5] \\
UHS J0259+3054 & [M6.5] & UHS J1200+2647 & [M4] \\
UHS J0305+5246 & [M8] & UHS J1259+5303 & [M6] \\
UHS J0319+0156 & [M8.5] & UHS J1304+4116 & [M8] \\
UHS J0321+2338 & [M8.5] & UHS J1359+1534 & [M7] \\
UHS J0322+1211 & [M6] & UHS J1359+5016 & [M6] \\
UHS J0341+0333 & [M6] & UHS J1415+3954 & [M6.5] \\
UHS J0345+0749 & [M6] & UHS J1421+3604 & [M5.5] \\
UHS J0403+0636 & [M8] & UHS J1423+2204 & [M5] \\
UHS J0447+1124 & [M6] & UHS J1426+4946 & [M8.5] \\
UHS J0452+1344 & [M6.5] & UHS J1429+2120 & [M8.5] \\
UHS J0531+2002 & [M3.5] & UHS J1459+1346 & [M8] \\
UHS J0538+5325 & [M3.5] & UHS J1503+1515 & [M7] \\
UHS J0540+4048 & [M4.5] & UHS J1517+5553 & [M8.5] \\
UHS J0542+5027 & [M3.5] & UHS J1545+2724 & [M6] \\
UHS J0551+4241 & [M6.5] & UHS J1607+0711 & [M6] \\
UHS J0555+5214 & [M5.5] & UHS J1705+0922 & [M6] \\
UHS J0603+4717 & [M6] & UHS J1749+2937 & [M7] \\
UHS J0607+4456 & [M5.5] & UHS J1752+3955 & [M4] \\
UHS J0611+0759 & [M8.5] & UHS J1759+0927 & [M5.5] \\
UHS J0621+5157 & [M8.5] & UHS J1812+1947 & [M6.5] \\
UHS J0646+1827 & [M6.5] & UHS J1816+1412 & [M4] \\
UHS J0655+4536 & [M6] & UHS J1824+0205 & [M5.5] \\
UHS J0710+5232 & [M4.5] & UHS J1824+5946 & [M7] \\
UHS J0711+1310 & [M7.5] & UHS J1837+4408 & [M6] \\
UHS J0711+1724 & [M6] & UHS J1841+2639 & [M4.5] \\
UHS J0717+2029 & [M5] & UHS J1850+2000 & [M6] \\
UHS J0722+0422 & [M5.5] & UHS J1910+2136 & [M5] \\
UHS J0722+4051 & [M7.5] & UHS J2002+1243 & [M4] \\
UHS J0723+1315 & [M5.5] & UHS J2028+1032 & [M5] \\
UHS J0723+3434 & [M5] & UHS J2052+1938 & [M6] \\
UHS J0726+2036 & [M6] & UHS J2135+1825 & [M9] \\
UHS J0729+3319 & [M6] & UHS J2137+0304 & [M5.5] \\
UHS J0755+0458 & [M6] & UHS J2200+0819 & [M5.5] \\
UHS J0855+5831 & [M5] \\
\enddata
\tablenotetext{*}{Spectral types were derived from the subdwarf sequence described in section \ref{sec:analysis}}
\end{deluxetable}

We note that the Gaia sample of discoveries from this search does not exactly match the subdwarf sequence (see Figure \ref{fig:cmd}), with some objects clearly being bluer and/or fainter (especially at M3 to M6 spectral types) and some objects redder/brighter than the subdwarf sequence (primarily at spectral types later than $\sim$M6). However, the subdwarf sequence does generally match the new candidates better than the empirical main sequence described by the GCNS. We therefore consider the spectral type estimates from this work preliminary, with follow-up spectroscopy necessary to determine accurate spectral types and metallicity class.

\begin{figure*}
\plotone{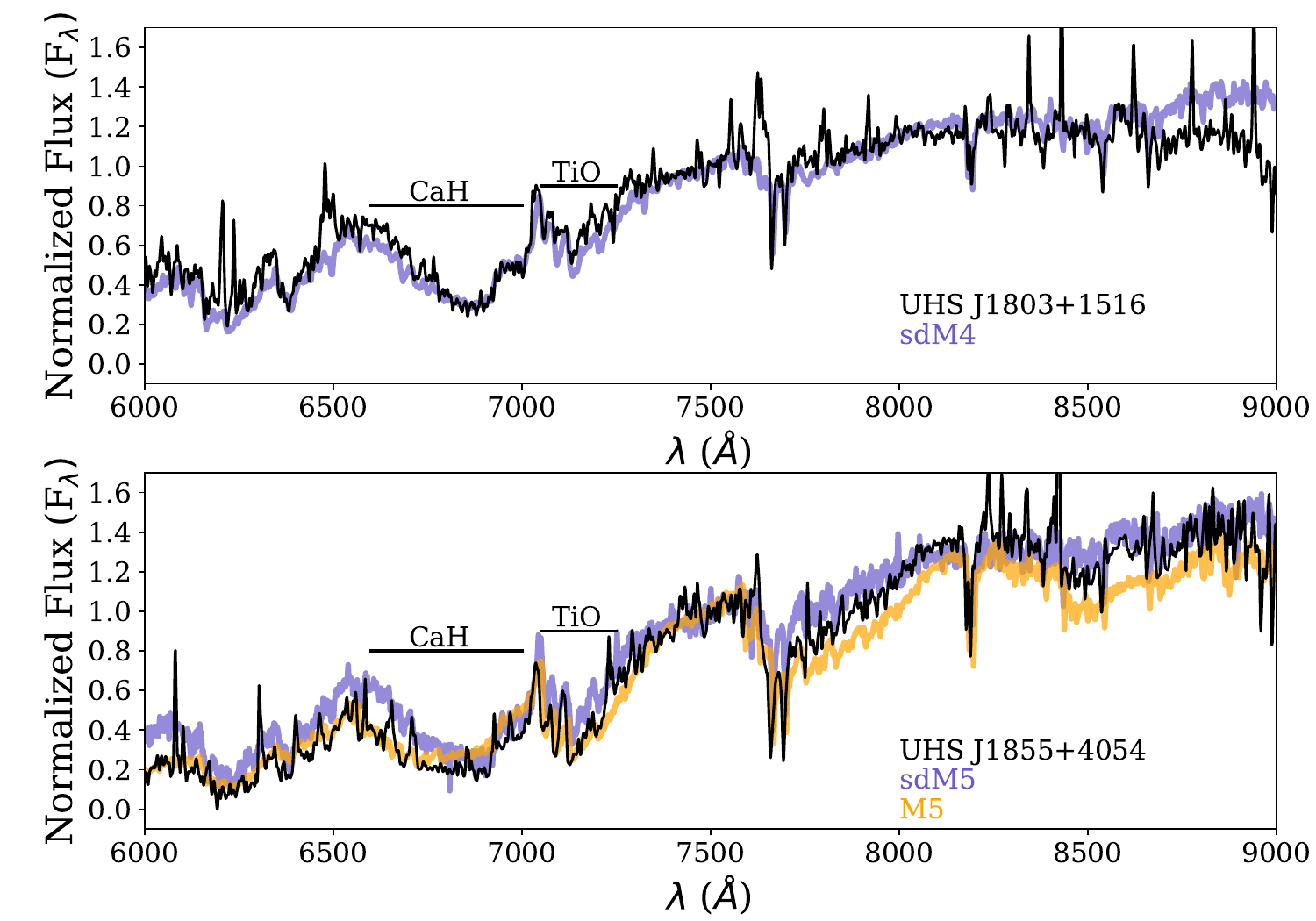}
\caption{Lick/KAST spectra (black) of J1803+1516 (top) and UHS J1855+4054 (bottom). The best fitting spectral standards from \cite{bochanski2007} and/or \cite{lepine2007} are shown as colored lines and labeled in each panel. All spectra are normalized at 7500\AA .}
\label{fig:spectra}
\end{figure*}

For two sources (UHS J1803+1516 and UHS J1855+4054), optical spectra were obtained as described in Section \ref{sec:obs} and are shown in Figure \ref{fig:spectra}. To determine spectral types for these sources, we compared each spectrum to optical M dwarf standards from \cite{bochanski2007} and \cite{lepine2007}. As seen in Figure \ref{fig:spectra}, UHS J1803+1516 is well-fit by the sdM4 standard, while UHS J1855+4054 shows a spectrum intermediate between the M5 and sdM5 standards. We therefore classify UHS J1803+1516 as sdM4 and UHS J1855+4054 as d/sdM5. To further analyze the spectral classifications of these objects, we calculate the $\zeta$-index as defined in \cite{lepine2007} with the updated coefficients in \cite{dhital2012}. The $\zeta$-index quantifies the stength of TiO and CaH molecular bands, as TiO band strengths weaken at lower metallicities. Stars with Solar metallicities have $\zeta$-index values around 1, while the most metal-poor stars have $\zeta$-index values of 0. Prominent TiO and CaH absorption regions are highlighted in Figure \ref{fig:spectra}. We find $\zeta$-index values of 0.53 and 0.87 for UHS J1803+1516 and UHS J1855+4054, respectively. The $\zeta$-index value of 0.53 for UHS J1803+1516 firmly supports its subdwarf classification, while the $\zeta$-index value of 0.87 is consistent with this object being near the boundary of $\zeta$ = 0.825 between dwarf and subdwarf M-type stars.

\subsection{Non-Gaia Sources} 
Our search uncovered 33 high-proper motion sources without counterparts in Gaia DR3, with 3 additional objects in Gaia DR3 but without parallax or proper motion measurements. To determine spectral type estimates for these sources, we follow the procedure detailed in the appendix of \cite{schneider2016}. This procedure employs a machine learning k-NN approach using available near-infrared $J$-, $H$-, and $K$-band photometry in combination with WISE W1, W2 and Pan-STARRS $grizy$ photometry, when available. All available colors are compared to a well-characterized training set of M, L, T, and Y dwarfs. Non-Gaia spectral type estimates are given in Table \ref{tab:ng_spt}. We find 10 likely M dwarfs, 15 objects with likely L spectral types, and 9 objects with T-type estimates. Sixteen new objects were found to have spectral types $\geq$L4, and are considered to be new brown dwarf candidates. This includes seven objects with L- spectral type estimates and nine early- to mid-T dwarfs. The latest spectral type estimate is T5 for UHS J0934+5037.

As with the Gaia sample, we determined $V_{\rm tan}$ estimates for these objects. First, we estimate distances for each non-Gaia source. For those with spectral type estimates L0 or later, we use apparent J-band magnitudes provided by UHS and the $M_{\rm J}$ vs.\ spectral type relation in Table 5 of \cite{schneider2023}. For those with spectral type estimates from M6 to M9, we use the $M_{\rm J}$ vs.\ spectral type relation from \cite{dupuy2012}. For the two objects with M5 spectral type estimates, which have G magnitudes in Gaia DR3 but no parallax measurements, we use the subdwarf sequence in Table \ref{tab:sds} to estimate distances. Note that these distance estimates have large uncertainties due to inexact spectral type estimates, uncertainties in the relations themselves, lack of binarity information, and low-S/N photometry (in some cases). From the distance estimates and UHS DR3 proper motions, we estimate $V_{\rm tan}$ values for each source. Uncertainties are handled in a Monte Carlo fashion. Distance and $V_{\rm tan}$ estimates are given in Table \ref{tab:ng_spt}.

The object estimated to be the closest among the sample is UHS J0023+4353, with spectral type estimate of L9 and a distance estimate of $\sim$36 pc (note that this object was also identified as a high proper motion source in \citealt{karpov2025}). The furthest distance estimates belong to the two objects with M5 spectral type estimates (UHS J0423+0016; $\sim$638 pc, UHS J0440+0831; $\sim$682 pc).

Perhaps unsurprising in a list of high proper motion discoveries, the tangential velocity estimates for this sample are large, with the smallest value belonging to the (likely) nearest source UHS J0023+4353 ($V_{\rm tan}$ $\sim$59$\pm$11 km s$^{-1}$). None of these objects satisfy the $<$10\% probability of thin disk membership criteria of \cite{dupuy2012} as a result of their large uncertainties. The likelihood of halo or thick disk membership is still high for many of these sources, with confirmed spectral types or refined distance measurements needed to better understand their nature. Several of these objects have tangential velocity estimates larger than the escape velocity of the Milky Way ($\approx$500--600 km s$^{-1}$; \citealt{williams2017, monari2018, koppelman2021}), highlighting the need for follow-up observations to further investigate their origins. 

\begin{figure}
\plotone{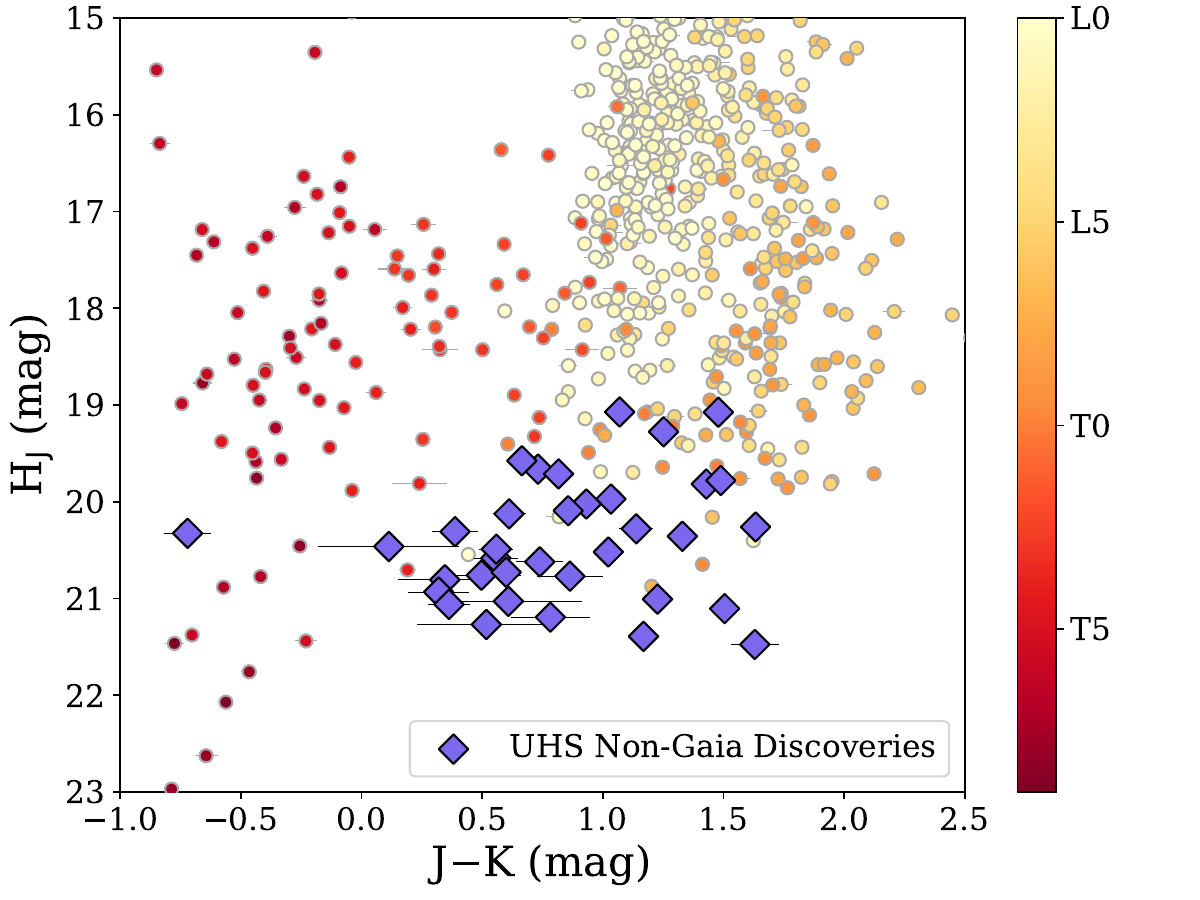}
\caption{Reduced proper motion diagram comparing the non-Gaia discoveries from this work to the sample of known L and T dwarfs within the UHS survey from \cite{schneider2023}.}
\label{fig:rpm}
\end{figure}

Figure \ref{fig:rpm} shows a reduced proper motion diagram that includes the $J$-band reduced proper motion ($H_{\rm J}$) versus $J-K$ color for our sample of non-Gaia discoveries compared to the sample of known L and T dwarfs detected in the UHS footprint from \cite{schneider2023}. The $J$-band reduced proper motion is calculated as H$_{J}$ = $J+$5log$_{10}$($\mu$)$+5$, where $\mu$ is the total proper motion in arcseconds. On reduced proper motion diagrams, old objects with high kinematics generally have larger $H_{\rm J}$ values, therefore falling lower on the figure. The placement of the sample of new discoveries from this work on the reduced proper motion diagram in Figure \ref{fig:rpm} supports the assertion that many of these objects likely have atypical kinematics. 

\begin{deluxetable*}{ccccc}
\tabletypesize{\scriptsize}
\tablecaption{Properties of Non-Gaia Sources}
\label{tab:ng_spt}
\tablehead{\colhead{UHS} & \colhead{Spectral} & \colhead{$d$\tablenotemark{a}} & \colhead{$V_{\rm tan}$\tablenotemark{a}} \\
\colhead{Name} & \colhead{Type} & \colhead{(pc)} & \colhead{(km s$^{-1}$)}}
\startdata
UHS J0023+4353	&	[L9]	&	[36$\pm$7]	&	[59$\pm$11]	\\
UHS J0043+2754	&	[L3]	&	[61$\pm$13]	&	[92$\pm$19]	\\
UHS J0047+3111	&	[L1]	&	[128$\pm$28]	&	[183$\pm$39]	\\
UHS J0209+3311	&	[T4]	&	[60$\pm$11]	&	[97$\pm$19]	\\
UHS J0257+1412	&	[T1]	&	[44$\pm$8]	&	[64$\pm$12]	\\
UHS J0307+3506	&	[M7]	&	[271$\pm$62]	&	[522$\pm$120]	\\
UHS J0423+0016	&	[M5]	&	[638$\pm$99]	&	[1035$\pm$163]	\\
UHS J0440+0831	&	[M5]	&	[682$\pm$106]	&	[1271$\pm$200]	\\
UHS J0449+0919	&	[T1]	&	[55$\pm$10]	&	[86$\pm$16]	\\
UHS J0609+3539  &   [M8]    &   [173$\pm$34]  & [467$\pm$91]  \\
UHS J0727+4003	&	[M8]	&	[195$\pm$38]	&	[297$\pm$58]	\\
UHS J0806+0309	&	[M6]	&	[493$\pm$157]	&	[844$\pm$268]	\\
UHS J0922+1306  &   [M8]    &   [203$\pm$40]  & [391$\pm$77] \\
UHS J0934+5037	&	[T5]	&	[46$\pm$9]	&	[67$\pm$13]	\\
UHS J1036+3140	&	[L5]	&	[67$\pm$13]	&	[107$\pm$22]	\\
UHS J1101+4307	&	[L6]	&	[52$\pm$10]	&	[76$\pm$15]	\\
UHS J1129+3704	&	[L8]	&	[76$\pm$15]	&	[135$\pm$27]	\\
UHS J1201+1632	&	[M8]	&	[177$\pm$35]	&	[251$\pm$49]	\\
UHS J1208+1720	&	[T2]	&	[71$\pm$14]	&	[109$\pm$22]	\\
UHS J1326+5543	&	[M8]	&	[164$\pm$32]	&	[257$\pm$51]	\\
UHS J1408+2140	&	[T3]	&	[63$\pm$12]	&	[94$\pm$18]	\\
UHS J1413+3121	&	[M6]	&	[439$\pm$81]	&	[644$\pm$120]	\\
UHS J1428+5214	&	[L0]	&	[226$\pm$53]	&	[405$\pm$95]	\\
UHS J1440+3528	&	[T3]	&	[69$\pm$13]	&	[100$\pm$19]	\\
UHS J1443+3119	&	[L6]	&	[73$\pm$15]	&	[139$\pm$28]	\\
UHS J1557+1212	&	[L2]	&	[72$\pm$15]	&	[193$\pm$40]	\\
UHS J1621+4318	&	[L0]	&	[158$\pm$36]	&	[243$\pm$56]	\\
UHS J1736+0543	&	[L5]	&	[43$\pm$9]	&	[62$\pm$12]	\\
UHS J1847+2905	&	[M7]	&	[318$\pm$73]	&	[503$\pm$117]	\\
UHS J1948+1112	&	[T0]	&	[49$\pm$9]	&	[72$\pm$14]	\\
UHS J2125+2142	&	[L4]	&	[83$\pm$17]	&	[125$\pm$25]	\\
UHS J2147+1627	&	[L1]	&	[95$\pm$21]	&	[152$\pm$33]	\\
UHS J2216+3136	&	[L2]	&	[53$\pm$11]	&	[99$\pm$21]	\\
UHS J2244+4655	&	[L0]	&	[160$\pm$37]	&	[293$\pm$68]	\\
UHS J2302+3254	&	[M7]	&	[248$\pm$57]	&	[415$\pm$95]	\\
UHS J2341+1939	&	[T0]	&	[52$\pm$10]	&	[117$\pm$22]	\\
\enddata
\tablenotetext{a}{Square brackets on the distance and tangential velocity values indicate that these estimates are based on photometric spectral types. }
\end{deluxetable*}

\section{Conclusion}
\label{sec:conclusion}
We examined the third data release of the UKIRT Hemisphere Survey for any previously unknown high proper motion objects with total proper motion $\gtrsim$300 mas yr$^{-1}$. This resulted in a total of 127 new high proper motion objects, including 16 brown dwarf candidates with spectral type estimates $\geq$L4. Using available photometry, the spectral types of all candidates were estimated. Aside from the L and T dwarf candidates, most of the new proper motion candidates are low mass and likely low-metallicity M-type stars, generally falling below the main sequence on color-magnitude diagrams, suggesting a subdwarf nature. The placement on the color-magnitude diagram suggests that some of the candidates may be even older and more metal-poor, which makes them potential extreme- or ultra-subdwarfs. Moreover, many of our candidates have interesting kinematics and are likely members of the galactic thick disk or halo. Optical spectroscopy of two sources confirms their low-mass and low-metallicity nature, with spectral types of sdM4 and d/sdM5 for UHS J1803+1516 and UHS J1855+4054, respectively. 

For future studies, follow-up spectroscopy of the candidates is necessary to confirm the spectral types and further characterize them. Spectra for some sources may soon be available via public releases from the Euclid Space Telescope \citep{laureijs2011} and/or the Spectro-Photometer for the History of the Universe, Epoch of Reionization, and Ices Explorer (SPHEREx; \citealt{dore2014}). Radial velocity measurements and distance measurements for the non-Gaia sources would allow for a deeper kinematic study of this sample. Additionally, future UHS searches for proper motion objects outside of the limits of this search, $\lesssim$300 mas yr$^{-1}$ or $\gtrsim$1200 mas yr$^{-1}$, could lead to more interesting discoveries and greatly expand our archive of proper motion objects. 

\begin{acknowledgments}
This publication makes use of data products from the UKIRT Hemisphere Survey, which is a joint project of the United States Naval Observatory, the University of Hawaii Institute for Astronomy, the Cambridge University Cambridge Astronomy Survey Unit, and the University of Edinburgh Wide-Field Astronomy Unit (WFAU). The WFAU gratefully acknowledges support for this work from the Science and Technology Facilities Council (STFC) through ST/T002956/1 and previous grants. SD acknowledges support from a UK STFC grant (ST/X000982/1).

This research made use of the SIMBAD database, operated at CDS, Strasbourg, France.

The Pan-STARRS1 Surveys (PS1) and the PS1 public science archive have been made possible through contributions by the Institute for Astronomy, the University of Hawaii, the Pan-STARRS Project Office, the Max-Planck Society and its participating institutes, the Max Planck Institute for Astronomy, Heidelberg and the Max Planck Institute for Extraterrestrial Physics, Garching, The Johns Hopkins University, Durham University, the University of Edinburgh, the Queen's University Belfast, the Harvard-Smithsonian Center for Astrophysics, the Las Cumbres Observatory Global Telescope Network Incorporated, the National Central University of Taiwan, the Space Telescope Science Institute, the National Aeronautics and Space Administration under Grant No. NNX08AR22G issued through the Planetary Science Division of the NASA Science Mission Directorate, the National Science Foundation Grant No. AST–1238877, the University of Maryland, Eotvos Lorand University (ELTE), the Los Alamos National Laboratory, and the Gordon and Betty Moore Foundation.

This publication makes use of data products from the Wide-field Infrared Survey Explorer, which is a joint project of the University of California, Los Angeles, and the Jet Propulsion Laboratory/California Institute of Technology, funded by the National Aeronautics and Space Administration. This publication also makes use of data products from NEOWISE, which is a project of the Jet Propulsion Laboratory/California Institute of Technology, funded by the Planetary Science Division of the National Aeronautics and Space Administration.

The authors wish to recognize and acknowledge the very significant cultural role and reverence that the summit of Mauna Kea has always had within the indigenous Hawaiian community. We are extremely grateful to have the opportunity to conduct observations from this mountain. 

\end{acknowledgments}

\facilities{UKIRT, Lick/Shane}

\software{\texttt{kastredux}}

\bibliography{references1}{}
\bibliographystyle{aasjournal}

\end{document}